\documentstyle[12pt]{article}
\begin{document}
\pagestyle{empty}
\vspace* {13mm}
\renewcommand{\thefootnote}{\fnsymbol{footnote}}
\begin{center}
   {\bf QUANTUM GROUPS, CORRELATION FUNCTIONS AND \\
	INFRARED DIVERGENCES} \\[25MM]
   Haye Hinrichsen and Vladimir Rittenberg \\[5mm]
   {\it Universit\"{a}t Bonn,
   Physikalisches Institut \\ Nussallee 12,
   D-5300 Bonn 1, Germany}
\\[5cm]
{\bf Abstract}
\end{center}
\renewcommand{\thefootnote}{\arabic{footnote}}
\addtocounter{footnote}{-1}
\vspace*{2mm}
We show in two simple examples that for
one-dimensional quantum chains with quantum group
symmetries, the correlation functions of local
operators are, in general, infrared divergent.
If one considers, however, correlation functions
invariant under the quantum group, the divergences
cancel out.
\vspace{4cm}
\begin{flushleft}
   BONN HE-93-02 \\
   hep-th/9301056 \\
   January 1993 \\
\end{flushleft}
\thispagestyle{empty}
\mbox{}
\newpage
\setcounter{page}{1}
\pagestyle{plain}
\noindent
In this paper we consider the anisotropic $XY$ chain
in a magnetic field which is defined by the Hamiltonian
\begin{equation}
\label{TwoParameterHamiltonian}
H(q,\eta) \; = \;
-\frac{1}{2}\sum_{j=1}^{L-1}\:
\left( \,\eta\:\sigma_{j}^{x}\sigma_{j+1}^{x}
\; + \; \eta^{-1}\sigma_{j}^{y}\sigma_{j+1}^{y}
\; + \; q\,\sigma_{j}^{z} \; + \;
q^{-1}\sigma_{j+1}^{z} \right) \;,
\end{equation}
\noindent
where $q$ and $\eta$ are complex parameters
and $\sigma^{x,y,z}_j$
are Pauli matrices placed on site $j$.
Up to boundary terms, which will play a
crucial role in our discussion, $H$ can be
rewritten as
\begin{equation}
\label{PeriodicHamiltonian}
H \; = \;
-\frac{1}{2}\sum_{j=1}^{L}\:
\left( \,\eta\:\sigma_{j}^{x}\sigma_{j+1}^{x}
\; + \; \eta^{-1}\sigma_{j}^{y}\sigma_{j+1}^{y}
\right)
\;-\; h\, \sum_{j=1}^{L}\; \sigma_j^z
\end{equation}
with $h=\frac{q+q^{-1}}{2}$. This Hamiltonian
has a long history \cite{Nijs} and provides
a good model for Helium adsorbed on metallic
surfaces ($\eta$ real and $q$ on the unit circle).
It also gives the master equation of the kinetic
Ising model \cite{Siggia} ($q=1$ and $\eta$ real).
The correlation functions
for this model given by Eq. (\ref{PeriodicHamiltonian})
with periodic boundary conditions
($\sigma^{x,y,z}_{L+1}=\sigma^{x,y,z}_1$)
have been computed for $\eta$ and $h$ real
\cite{McCoy}. The calculations are relatively simple
since the Hamiltonian (\ref{PeriodicHamiltonian})
can be diagonalized in terms of free fermions.
It is the aim of this paper to compute the
correlation functions of the Hamiltonian given
by Eq. (\ref{TwoParameterHamiltonian}) in
the massless regime. We will be concerned with
two cases: $\eta=1$ and $q$ on the unit circle
(notice that in this case $H$ given by Eq.
(\ref{TwoParameterHamiltonian}) is not
hermitian whereas $H$ in Eq.
(\ref{PeriodicHamiltonian}) is) and $q=1$,
$\eta$ on the unit circle (where $H$ in
Eqs.~(\ref{TwoParameterHamiltonian})
and (\ref{PeriodicHamiltonian}) are
nonhermitian). Notice that the correlation
functions in the latter case have not yet
been computed with any boundary conditions.
Our interest in these
calculations goes beyond the particular
physics of the model.
As opposed to $H$ defined by
Eq. (\ref{PeriodicHamiltonian}), the
Hamiltonian defined by Eq.
(\ref{TwoParameterHamiltonian}) is invariant
under quantum group deformations, and we
want to use the fact that $H$ can be
easily diagonalized and the correlation
functions can be computed in order to find
out if there are unexpected phenomena
in the massless regime (it turns out that
there are). For other quantum chains with higher
rank quantum symmetries or supersymmetries, the
calculations are much more involved. \\
\indent
In order to see the symmetry properties
\cite{Ourself}
we first perform a Jordan-Wigner
transformation and introduce new matrices
$\tau_j^1$ and $\tau_j^2$ by
\begin{equation}
\label{JordanWigner}
\tau_{j}^{1}\;=\;
  \Bigl(\prod_{i<j}
  \sigma_{i}^{z}\Bigr)\,\sigma_{j}^{x}
\;,\;\;\;\;\;\;\;\;\;
\tau_{j}^{2}\;=\;\Bigl(\prod_{i<j}
  \sigma_{i}^{z}\Bigr)\,\sigma_{j}^{y}
\;\;\;\;\;\;\;\;\;\;\;\;\;\;\;\;\;\;\;\;\;\;\;\;\;
(j=1 \ldots L)
\end{equation}
\noindent
which obey the Clifford algebra
\begin{equation}
\label{TauCommutators}
\{\tau_{i}^{\mu},\tau_{j}^{\nu}\} \;=\;
2\ \delta_{i,j}^{\mu,\nu} \;
\;\;\;\;\;\;\;\;\;\;\;\;\;\;\;\;\;\;\;\;
(i,j=1 \ldots L;\;\;\mu,\nu=1,2)
\end{equation}
\noindent
and get
\begin{equation}
\label{TauHamiltonian}
H(q,\eta)\;=\; \frac{i}{2}\sum_{j=1}^{L-1}
\left(
   \eta\,\tau_{j}^{2}\tau_{j+1}^{1}
   \ - \  \eta^{-1}\,\tau_{j}^{1}\tau_{j+1}^{2}
   \ + \ q\,\tau_{j}^{1}\tau_{j}^{2}
   \ + \ q^{-1}\,\tau_{j+1}^{1}\tau_{j+1}^{2} \,
\right).
\end{equation}
\noindent
This Hamiltonian is invariant under
the two-parameter ($\alpha$ and $\beta$)
quantum algebra
\begin{eqnarray}
\label{TwoParamAlgebra}
 & \{T_0^1,T_0^1\} \;=\; 2\:[E]_{\alpha} \ \ \ \ \ \ \ \
 & \{T_0^1,T_0^2\} \;=\; 0  \\
 & \{T_0^2,T_0^2\} \;=\; 2\:[E]_{\beta} \ \ \ \ \ \ \ \ \
 & [E,T_0^1] \;=\; [E,T_0^2] \;=\; 0\,\nonumber
\end{eqnarray}
\noindent
together with a suitable coproduct
\cite{Ourself}, where
\begin{equation}
\label{AlphaBeta} \alpha=\frac q\eta \,,
\hspace{1cm}
\beta=q\eta \,,
\hspace{1.5cm}
[E]_{\alpha} \;=\;
 \frac{\alpha^{E}-\alpha^{-E}}
{\alpha-\alpha^{-1}}\;.
\end{equation}
\noindent
In the representation of the quantum chain
(\ref{TauHamiltonian}), the operators $T_0^1$, $T_0^2$,
and $E$ are obtained from the coproduct rules
\cite{Ourself} and read
\begin{equation}
\label{ZeroModeOperators}
T_0^{1} \;=\;
    \alpha^{-\frac{L+1}{2}} \;\sum_{j=1}^{L}
    \alpha^{j} \,\tau_{j}^{1} \,,
\hspace{1.5cm}
T_0^{2} \;=\;
    \beta^{-\frac{L+1}{2}} \;\sum_{j=1}^{L}
    \beta^{j} \,\tau_{j}^{2} \,,
\hspace{1.5cm}
E\;=\;L\,,
\end{equation}
\noindent
where $L$ is the length of the chain.
For generic $\alpha$ and $\beta$
the nontrivial irreducible representations
of the algebra (\ref{TwoParamAlgebra}) are two-dimensional.
Because of the branching rules given by the representation
theory of the algebra (\ref{TwoParamAlgebra})
all energy levels of $H$ are twice degenerated.\\
\indent
If $\eta=1$,
the total magnetization operator
$N=\frac12\sum_{j=1}^L\sigma_j^z=
\frac i2\sum_{j=1}^L \tau_j^2\tau_j^1$
also commutes with the Hamiltonian. In this case
the algebra (\ref{TwoParamAlgebra})
can be enlarged by
the commutation relations
\begin{equation}
\label{Additional}
[N,T_0^1] \;=\; i \, T_0^2
\hspace{1.5cm}
[N,T_0^2] \;=\; - i \, T_0^1
\hspace{1.5cm}
[N,E]\;=\;0\;.
\end{equation}

\noindent
The coproduct for the operator $N$
can be read off from its definition. This is
the quantum superalgebra $U_qSU(1|1)$
\cite{Saleur}. The Hamiltonian
(\ref{TwoParameterHamiltonian}) is also obtained
if one asks for $SU(2)_q$ symmetry for $q=i$
and one uses unrestricted
representations \cite{Daniel}.
\\
\indent
Before computing the correlation functions, we
give the diagonalized form of $H$ \cite{Ourself}
\begin{equation}
\label{DiagonalForm}
H(q,\eta)\;=\;\sum_{k=1}^{L-1}\,
\Lambda_{k}\;i T_{k}^{2}T_{k}^{1} \;,
\end{equation}
\noindent
where
\begin{equation}
\label{Dispersion}
\Lambda_k \;=\;
\left(
\frac{1}{4\alpha\beta} \,
\rm
(\alpha-e^{\frac{i \pi k}{L}})\,
(\alpha-e^{-\frac{i \pi k}{L}})\,
(\beta-e^{\frac{i \pi k}{L}})\,
(\beta-e^{-\frac{i \pi k}{L}})
\right)^\frac12
\end{equation}
are fermionic excitation energies.
The operators $T_k^\gamma$ are certain
linear combinations of the $\tau_j^\mu$
and obey the commutation
relations
\begin{equation}
\label{Clifford2}
\{T_k^\mu,T_l^\nu\} = 2 \,\delta_{k,l}^{\mu,\nu} \;,
\hspace{1.3cm}
\{T_0^\mu,T_k^\nu\}=0 \;.
\hspace{2cm}
(k,l=1\ldots L-1; \; \mu,\nu=1,2)
\end{equation}
This implies that the operators
$iT_k^2T_k^1$ have the
eigenvalues $\pm1$ and commute for
different~$k$.
For complex parameters $q$ and $\eta$
the operators $T_k^\mu$ are
not hermitian. Notice that $T_0^1$ and
$T_0^2$ do not appear in $H$. Observe also
(see Eq. (\ref{Dispersion})), that
the spectrum of $H(q,\eta)$ coincides
with the spectrum of $H(\eta,q)$
(this is a generalized duality property
\cite{Ourself}).
\\
\indent
As mentioned above the ground state
is twofold degenerate.
The corresponding subspace
is spanned by the basis vectors
$|0^+\rangle$ and $|0^-\rangle$ which are
defined by
\begin{eqnarray}
\label{Annihilate}
&& i\,T_k^2T_k^1\,|0^\pm\rangle \;=\;
                 -|0^\pm\rangle
\\
&& i\,T_0^2T_0^1\,|0^\pm\rangle \;=\;
      \pm \sqrt{[L]_\alpha [L]_\beta}\;\,
      |0^\pm\rangle \,
\nonumber
\end{eqnarray}
\noindent
where $1 \leq k \leq L-1$.
The corresponding bra-vectors are defined by
\begin{eqnarray}
\label{Annihilate2}
&& \langle 0^\pm| \,i\,T_k^2T_k^1 \;=\;
                 - \langle 0^\pm |
\\
&& \langle 0^\pm|\,i\,T_0^2T_0^1 \;=\;
     \pm \sqrt{[L]_\alpha [L]_\beta}\;\,
     \langle 0^\pm | \, \nonumber
\end{eqnarray}
\noindent
Although in general the Hamiltonian
(\ref{TwoParameterHamiltonian})
is non-hermitian,
it is symmetric, i.e. $H=H^T$.
Thus going from the
bras to the kets we take the
transpose only (no complex conjugation).
\\
\indent
Since $T_0^\mu\,|0^\pm\rangle \neq 0$, it is
impossible to define a linear combination of
$|0^+\rangle$ and $|0^-\rangle$ as
a scalar with respect to the quantum algebra. In
the special case $\eta=1$, where
the algebra can be enlarged by the
relations (\ref{Additional}),
the states $|0^+\rangle$ and $|0^-\rangle$ are
also eigenstates of the additional operator $N$.
Thus we have to choose either
$|0^+\rangle$ or $|0^-\rangle$ as the ground
state of the system. \\
\indent
Now we consider the
correlation function
\begin{equation}
\label{ZZCorr}
G^{zz}(R,S,L)=
\langle 0^\pm|\sigma_l^z\sigma_m^z|0^\pm\rangle =
\langle 0^\pm|i\tau^2_l\tau^1_l|0^\pm\rangle
\langle 0^\pm|i\tau^2_m\tau^1_m|0^\pm\rangle -
\langle 0^\pm|i\tau^2_l\tau^1_m|0^\pm\rangle
\langle 0^\pm|i\tau^2_m\tau^1_l|0^\pm\rangle \,,
\end{equation}
\noindent
where we have used the Wick theorem and
taken $R=m-l$ and
$S=m+l$. One can show that
\begin{eqnarray}
\label{ExactExpression1}
\langle 0^\pm| i\tau^2_m\tau^1_l |0^\pm\rangle & = &
\pm\frac{q^{S-L-1}\eta^{R}}{\sqrt{[L]_\alpha[L]_\beta}}
 \; + \;
\frac{1}{4Lq}
  \sum^{L-1}_{
  \begin{array}{c}
  \\[-7mm]
  {\scriptstyle k=-L+1} \\[-2.2mm]
  {\scriptstyle k \neq 0}
  \end{array}}
\Lambda_k^{-1}\; \times \\
&& \Bigl[\, (\beta-e^{-\frac{i \pi k}{L}})
      (\alpha-e^{\frac{i \pi k}{L}})
      e^{\frac{i \pi k}{L} R}
      \,-\,
      (\beta-e^{-\frac{i \pi k}{L}})
      (\alpha-e^{-\frac{i \pi k}{L}})
      e^{\frac{i \pi k}{L} S}
\;\Bigr]\;. \nonumber
\end{eqnarray}

\noindent
We want to compute the
limit $G^{zz}(R,S) =
\lim_{L \rightarrow \infty} G^{zz}(R,S,L)$
with $R$ and $S$ fixed.
If the spectrum is massive, the
factor $\Lambda_k^{-1}$ in Eq.
(\ref{ExactExpression1}) is bounded and we
can replace the sum by an integral. This
integral is finite and thus the limit
$L \rightarrow \infty$ exists in the
massive case. We are interested however in the
massless case where the situation is different.\\
\indent
By Eq. (\ref{Dispersion})
the spectrum is massless if at least one of the
parameters $\alpha$ or $\beta$ lies on the unit
circle. Hence in terms of $q$ and $\eta$, there
are three massless situations:
$(q \in S^1, \eta=1)$, $(q=1, \eta \in S^1)$
and $(q \in S^1, \eta \in S^1)$. In the last
case the spectrum
contains imaginary excitations
and the states $|0^\pm\rangle$ defined
in Eq. (\ref{Annihilate}) loose their
physical meaning as states of minimal
energy. This case will not be considered
(The Ising line $q=\eta$, also massless, is not
interesting for the present investigation).
\\
\indent
Let us consider the first case where
\underline {$\eta=1$ and $q=e^{i \pi \phi}$}.
We would expect the thermodynamic limit
$L \rightarrow \infty$
of correlation
functions to exist even for
non-periodic boundary conditions in both the
massive and the massless case. For example
the correlation function for a conformally
invariant system with free boundary conditions
on a strip
\cite{Cardy} has the form
\begin{equation}
\label{CardyForm}
\lim_{L \rightarrow \infty}
G^{strip}(R,S,L) \;=\;
G^{strip}(R,S) \;=\;
\frac1{R^{2x}}\,F(S^2/R^2)\,,
\end{equation}
\noindent
where $x$ is the bulk exponent. Here $F$ is
a function which satisfies
$F(\rho) \stackrel {\rho \rightarrow \infty}
{=} 1$ giving the correlation function in the bulk
\begin{equation}
\label{Bulk}
G^{strip}(R) \;=\; \frac1{R^{2x}}\,.
\end{equation}
On the other hand
$F(1+\delta) \stackrel {\delta \rightarrow 0}
{=} \delta^{x_s-x}$, where
$x_s$ is the surface exponent. This gives the
correlation function when the first point
(in the notation of
Eq. (\ref{ZZCorr}) at position $l$)
is kept near the boundary
and the second is
taken far away:
\begin{equation}
\label{Surface}
G^{strip}(l,R) \;=\; \frac{(4l)^{x_s-x}}{R^{x_s+x}}\,.
\end{equation}
\\
\indent
For the boundary conditions defined in Eq.
(\ref{TwoParameterHamiltonian}), where
$H$ is non-hermitian in the first
and the last position of the chain, we will
find however divergent results
as we will show in the following.
If $q=e^{i \pi \phi}=e^{i \pi \frac r s}$
and $L$ is a multiple
of $s$, Eq. (\ref{ExactExpression1})
is not defined since one of the
$\Lambda_k$ vanishes. Hence in
order to perform a thermodynamic limit,
we have to use a sequence in $L$ given by
$L=L's+t$, where $0<t<s$ and $L'$ are integers.
But even then one can show that  Eq.
(\ref{ExactExpression1})
diverges like $\log L$ when
$L' \rightarrow \infty$:
\begin{eqnarray}
\label{Divergence}
G^{zz}(R,S,L)
\;=\; & \log L & \biggl[ \,
\frac {4 e^{2 i \pi \phi (S-1)}
\sin^2 \pi \phi } {\pi^2}
\Bigl(
2 \log S - \log (S+R) - \log (S-R)
\Bigr) \\
&&  \;\;
+\,\frac {2 e^{i \pi \phi (S-1)}
\sin \pi \phi } \pi
\, \Bigl(
\frac{4 \sin \pi \phi R}{\pi R} +
(2-4 \phi) \cos \pi \phi R
\Bigr) \, \biggr] \;+\; \ldots \nonumber
\end{eqnarray}

\noindent
This divergence is
a consequence of the
boundary conditions in Eq.
(\ref{TwoParameterHamiltonian}) since for
periodic boundary conditions
(no $U_qSU(1|1)$ symmetry !),
one obtains the non-divergent
result \cite{McCoy}
\begin{equation}
\label{NonDivergent}
G^{zz}(R) \;=\;
m_z^2-\frac{4 \sin^2 \pi \phi R}{\pi^2 R^2}
\end{equation}
\noindent
where $m_z^2$ is the magnetization per site whose
expression is given in Ref. \cite{McCoy}.
The occurence of oscillatory terms reflects the
fact that we are here in an incommensurate phase.
Notice
that for $q=\eta=1$ the divergence in Eq.
(\ref{Divergence}) disappears.
Similar divergences occur if we take $\eta=1$
and $q=e^{i \pi \phi}-\varepsilon$ with
small $\varepsilon$ or $\eta=1+\varepsilon$,
$q=e^{i \pi \phi}$ and
$\varepsilon \rightarrow 0$. \\
\indent
In the second case, \underline {$q=1$ and
$\eta=e^{i \pi \phi}$}, we also obtain
a divergent result:
\begin{eqnarray}
\label{Divergence3}
G^{zz}(R,S,L)
&=&  \log L \;
\frac {8 \sin \pi \phi } {\pi^2} \,
\biggl[ \,
\Bigl(
\log R + \log (2 \sin \pi \phi) + C
\Bigr) \, \sin \pi \phi
\nonumber \\
&&\hspace{2.7cm}
+\, (1-2 \phi) \cos \pi \phi +
\frac{2 \sin \bigl( \pi \phi (S-1) \bigr)
\, \sin \pi \phi R}{S-1}  \\
&&\hspace{2.7cm}
-\,
\frac{\sin \bigl( \pi \phi (S+R-1) \bigr) }{S+R-1} -
\frac{\sin \bigl( \pi \phi (S-R-1) \bigr) }{S-R-1}
\; \biggr] \;+\; \ldots \nonumber
\end{eqnarray}
Here $C$ is the Euler constant.
In this case a similar calculation with
periodic boundary
conditions (the Hamiltonian
is non-hermitian) gives again an
infrared divergent result. \\[5mm]
\indent
The main message of this paper is that
there is a way to find correlation functions
with a well defined thermodynamic limit
in the presence of a quantum group. Instead of
computing the correlation functions of two arbitrary
local operators, let us consider the invariant
quantity:
\begin{equation}
\label{Combination}
c_{l,m} \;=\;
q^{m-l}\,i\tau_l^2\tau_l^1 \;+\;
q^{l-m}\,i\tau_m^2\tau_m^1 \;-\;
\eta^{m-l}\,i\tau_l^2\tau_m^1 \;-\;
\eta^{l-m}\,i\tau_m^2\tau_l^1 \,
\end{equation}
\noindent
This operator
is a combination of two local operators
$i\tau_l^2\tau_l^1=\sigma_l^z$ and
two dislocation lines
($i\tau_l^2\tau_m^1$ and
 $i\tau_m^2\tau_l^1$, see Ref. \cite{Nijs}).
It is symmetric ($c_{l,m}=c_{m,l}$) and
commutes with the generators of
the quantum group:
\begin{eqnarray}
\label{Symmetry}
&&[c_{l,m}\,,T_0^1] = [c_{l,m}\,,T_0^2] = 0
\end{eqnarray}
\noindent
Moreover $c_{l,m}$ is the
unique combination which is invariant
(there is only one Casimir operator for the algebra
(\ref{TwoParamAlgebra})).
For \underline {$\eta=1$ and $q=e^{i \pi \phi}$}
we derive from Eq. (\ref{ExactExpression1}) that  the
corresponding correlation function is given by
\begin{eqnarray}
\label{ExactExpression3}
G^c(R,S,L)=
\langle 0^\pm| c_{l,m} | 0^\pm \rangle &=&
\frac 1L
  \sum^{L-1}_{
  \begin{array}{c}
  \\[-7mm]
  {\scriptstyle k=-L+1} \\[-2.2mm]
  {\scriptstyle k \neq 0}
  \end{array}}
\mbox{sgn} \, (|\phi|-|\frac kL|)  \;\times \\[-1mm]
&&\biggl[ \,
2\,
\frac {\sin \Bigl( \frac \pi 2 (\frac k L + \phi) \Bigr) \,
       \sin^2 \Bigl( \frac \pi 2 (\frac k L - \phi) R \Bigr) }
      {\sin \Bigl( \frac \pi 2 (\frac k L - \phi) \Bigr) }
\, e^{i \pi \frac k L (S-1)} \nonumber \\
&& \;\;\;+\; \cos(\pi \frac k L R) \;-\; \cos (\pi \phi R)
\;\biggr]\,. \nonumber
\end{eqnarray}
\noindent
Since $c_{l,m}$ is invariant under
the quantum group, the correlation function
(\ref{ExactExpression3}) does not depend
on the choice of the ground state.
As one sees the correlation function is
infrared finite. Because
of the special combination in Eq.
(\ref{Combination}) the zero of the
denominator in Eq. (\ref{ExactExpression3})
(which comes from the zero in the dispersion
relation) is cancelled by a zero in
the numerator. In the limit
$L \rightarrow \infty$ we obtain
\begin{eqnarray}
\label{ExactSolution}
G^c(R,S) &=& \frac 2 \pi \; \biggl[ \;
2 i \sin \pi \phi \, \sum_{j=1}^{|R|-1}
\left(\;e^{i \pi \phi j}\frac{
     \sin \Bigl(\pi \phi (S-j-1)\Bigr)}{S-j-1}-
     e^{-i \pi \phi j}\frac{
     \sin \Bigl(\pi \phi (S+j-1)\Bigr)}{S+j-1}\right)
     \nonumber \\
&&\;\;\;\;\;-\;e^{i \pi \phi (R-1)}\frac{
           \sin \Bigl(\pi \phi (S-R-1)\Bigr)}
          {S-R-1} \;-\;
     e^{i \pi \phi (1-R)}\frac{
           \sin \Bigl(\pi \phi (S+R-1)\Bigr)}
          {S+R-1} \\
&&\;\;\;\;\;+\;2\cos\pi\phi
\frac{\sin \Bigl(\pi\phi(S-1) \Bigr)}{S-1}
\;+\; \frac{2 \sin \pi\phi R}{R} \;\biggr]
\;+\; (2-4\phi)\cos \pi \phi R\nonumber \;,
\end{eqnarray}
\noindent
which is an exact result.
We can now derive the "bulk" and "surface" behaviour
(see Eqs. (\ref{Bulk}) and (\ref{Surface}))
\begin{eqnarray}
\label{BulkInvariant}
G^c(R) &=&
(2-4\phi)\cos{\pi \phi R} +
\frac{4 \sin{\pi \phi R}}{\pi R}
+ 0(R^{-2}) \\
G^c(l,R) &\approx&
(2-4\phi)\cos{\pi \phi R} +
\frac 2 \pi \sin \pi \phi \,
e^{i \pi \phi (R+2l-1)} \,
\log \frac {|R|}{4l}
\,.
\end{eqnarray}
\noindent
We note that the chain given by Eq.
(\ref{TwoParameterHamiltonian}) is the
simplest one of a series of quantum chains,
the so-called Perk-Schultz ones
\cite{Perk}, which have as symmetry the
$U_qSU(M|N)$ quantum superalgebras
\cite{SUMN}, and we expect to find
similar results in all cases. We also
suspect that our results will have
consequences on the calculations of
correlation functions performed directly
in the $L \rightarrow \infty$ limit
using the quantum affine symmetry
\cite{Japan}.
\\
\indent
In the second case where \underline
{$q=1$ and $\eta=e^{i \pi \phi}$} we are led to
\begin{eqnarray}
\label{ExactExpr2}
G^c(R,S,L) & = &
\frac 1 L
  \sum^{L-1}_{
  \begin{array}{c}
  \\[-7mm]
  {\scriptstyle k=-L+1} \\[-2.2mm]
  {\scriptstyle k \neq 0}
  \end{array}}
\mbox{sgn} \, (|\phi|-|\frac kL|)  \;\times \nonumber \\[-1mm]
&&\biggl[\,
2\,
\frac {\sin \Bigl( \frac \pi 2 (\frac k L + \phi) \Bigr)\,
       \sin^2 \Bigl( \frac \pi 2 (\frac k L - \phi) R \Bigr) }
      {\sin   \Bigl( \frac \pi 2 (\frac k L - \phi) \Bigr) }
\nonumber \\
&&\;\;\;+\; e^{i \pi \frac k L (S-1)}\, \Bigl(
 \cos(\pi \phi R) \;-\; \cos (\pi \frac k L R)
\, \Bigr)
\;\biggr]\;. \nonumber
\end{eqnarray}
\noindent
Here we get the infrared finite result
\begin{eqnarray}
\label{ExactSol2}
G^c(R,S) & = & \frac 2 \pi \;\;\biggl[\;
 \Bigl(2+\sum_{j=1}^{|R|-1}
 \frac{2j+1}{j(j+1)} \Bigr)
  \, \sin \pi\phi
\;-\; \sum_{j=1}^{|R|-1}
 \frac{\sin \Bigl(\pi\phi(2j+1)\Bigr)}{j(j+1)} \\
&&\;\;\;\;\;\;\;\;+\;
 \Bigl(\frac1{S-1}-\frac1{S+R-1} \Bigr)
 \sin \Bigl(\pi\phi(S+R-1)\Bigr)
 \nonumber \\
&&\;\;\;\;\;\;\;\;+\;
 \Bigl(\frac1{S-1}-\frac1{S-R-1} \Bigr)
 \sin \Bigl(\pi\phi(S-R-1)\Bigr) \;\,\biggr]
\;+\; (2-4\phi)\,\cos\pi\phi \nonumber
\end{eqnarray}
\noindent
which is also an exact expression.
The "bulk" and the "surface" limit turn out to
be given by
\begin{eqnarray}
\label{Approx}
G^c(R) & \approx &
(2-4\phi)\,\cos\pi\phi +
\frac 2 \pi \sin \pi \phi
\left(
\log(4\sin^2\pi\phi)+\log R^2 + 2C
\right) \\
G^c(l,R) & \approx &
(2-4\phi)\,\cos\pi\phi +
\frac 2 \pi \sin \pi \phi
\left(
\log(4\sin^2\pi\phi)+\log R^2 + 2C
\right) -
\frac {2 \sin \pi \phi (2l-1)} {\pi (2l-1)}
\nonumber
\end{eqnarray}
\noindent
where $C$ is the Euler constant. \\
\indent
We do not yet have an explanation for the
"miraculous" cancellations of divergent
quantities which occur when considering
invariant quantities. Probably these
quantities "know" to pick up only the
right states of the Hilbert space, and in
this subspace a scalar product can be defined
in spite of the nonhermiticity of the
Hamiltonian. A similar but completely unrelated
phenomenon occurs in $\sigma$-models and
their generalizations \cite{Elitzur},
where again only group (not quantum group!)
invariant quantities give infrared finite
correlation functions. \\[5mm]
\indent
One of us (V.R.) would like to thank the
University of Geneva and the Einstein Center
of the Weizmann Institute for their
hospitality. We would also like to thank
R. Flume, A. Ganchev,
C. Gomez, A. Schwimmer and
E. Sokatchev for useful discussions.

%%%%%%%%%%%%%%%%%%%%%%%%%%%%%%%%%%%%%%%%%%%%%%%%%%%%%%%%%%%

\end{document}